\documentclass{ws-procs961x669}            
\begin{document}
\title{Laue lenses:\\
Focusing optics for hard X/soft Gamma-ray Astronomy}

\author{L. Ferro$^*$, M. Moita, P. Rosati, R. Lolli, C. Guidorzi, F. Frontera}

\address{Department of Physics and Earth Science, University of Ferrara,\\
Via Giuseppe Saragat 1, Ferrara (FE), 44122, Italy\\
$^*$E-mail: frrlsi@unife.it\\
www.fst.unife.it}

\author{E. Virgilli, E. Caroli, N. Auricchio, J. B. Stephen, C. Labanti, F. Fuschino, R. Campana}

\address{INAF/OAS-Bologna, \\
Via Piero Gobetti 93/3, Bologna (BO), 40129, Italy\\
www.oas.inaf.it}

\author{C. Ferrari}

\address{Institute of Materials for Electronics and Magnetism (IMEM), CNR, \\
Parco Area delle Scienze 37/A, Parma (PR), 43124, Italy\\
www.imem.cnr.it}

\author{S. Squerzanti}

\address{INFN of Ferrara, \\
Via Giuseppe Saragat 1, Ferrara (FE), 44122, Italy \\
www.fe.infn.it}

\author{M. Pucci}

\address{National Institute of Optics (Istituto Nazionale di Ottica, CNR-INO), CNR, \\
Largo Enrico Fermi 6, Firenze (FI), 50125, Italy \\
www.ino.cnr.it}

\author{S. del Sordo, C. Gargano}

\address{IASF of Palermo, INAF, \\
Via Ugo La Malfa 153, Palermo (PA), 90146, Italy\\
www.ifc.inaf.it}

\begin{abstract}
Hard X-/soft Gamma-ray astronomy is a key field for the study of important astrophysical phenomena such as the electromagnetic counterparts of gravitational waves, gamma-ray bursts, black holes physics and many more. However, the spatial localization, imaging capabilities and sensitivity of the measurements are strongly limited for the energy range $>$70 keV due to the lack of focusing instruments operating in this energy band. A new generation of instruments suitable to focus hard X-/ soft Gamma-rays is necessary to shed light on the nature of astrophysical phenomena which are still unclear due to the limitations of current direct-viewing telescopes.
Laue lenses can be the answer to those needs. A Laue lens is an optical device consisting of a large number of properly oriented crystals which are capable, through Laue diffraction, of concentrating the radiation into the common Laue lens focus. In contrast with the grazing incidence telescopes commonly used for softer X-rays, the transmission configuration of the Laue lenses allows us to obtain a significant sensitive area even at energies of hundreds of keV.
At the University of Ferrara we are actively working on the modelization and construction of a broad-band Laue lens. In this work we will present the main concepts behind Laue lenses and the latest technological developments of the TRILL (Technological Readiness Increase for Laue Lenses) project, devoted to the advancement of the technological readiness of Laue lenses by developing the first prototype of a lens sector made of cylindrical bent crystals of Germanium. 
\end{abstract}


\bodymatter

\section{Laue lenses: a new way to look at the X-ray/Gamma sky}
\label{sect:laue}
Laue lenses are innovative X and Gamma-ray optics based on the Bragg’s law of diffraction in crystals\cite{Zachariasen}:
\begin{equation}
    2 d_{hkl}\sin \theta_B = n \frac{hc}{E}
\end{equation}

where $\theta_B$ is the Bragg's angle, which is the 
angle between the diffraction planes of the crystal and the diffracted beam, $\mathrm{d_{hkl}}$ is the 
inter-planar spacing of diffraction planes of the crystal with Miller indexes $\mathrm{(hkl)}$, $\mathrm{E}$ is the energy of the diffracted photon, 
$n$ is the diffraction order and  $\mathrm{hc}$ 
corresponds to 12.39 keV\text{\AA}.

The crystals can be arranged in such a way that the X-ray beam can be reflected by a thin, superficial layer of the crystal (Bragg geometry, or Reflection geometry) or it can cross the whole crystal and be transmitted over it (Laue geometry, or Transmission geometry).

Optics based on Bragg's law of diffraction in Laue configuration can be of great interest in the astrophysical context because (1) Bragg's diffraction is effective up to energies of few MeV and (2) the transmission configuration can be exploited to increase the effective area of the optics.  

\begin{figure}[t!]
    \centering
    \includegraphics[scale = 0.25]{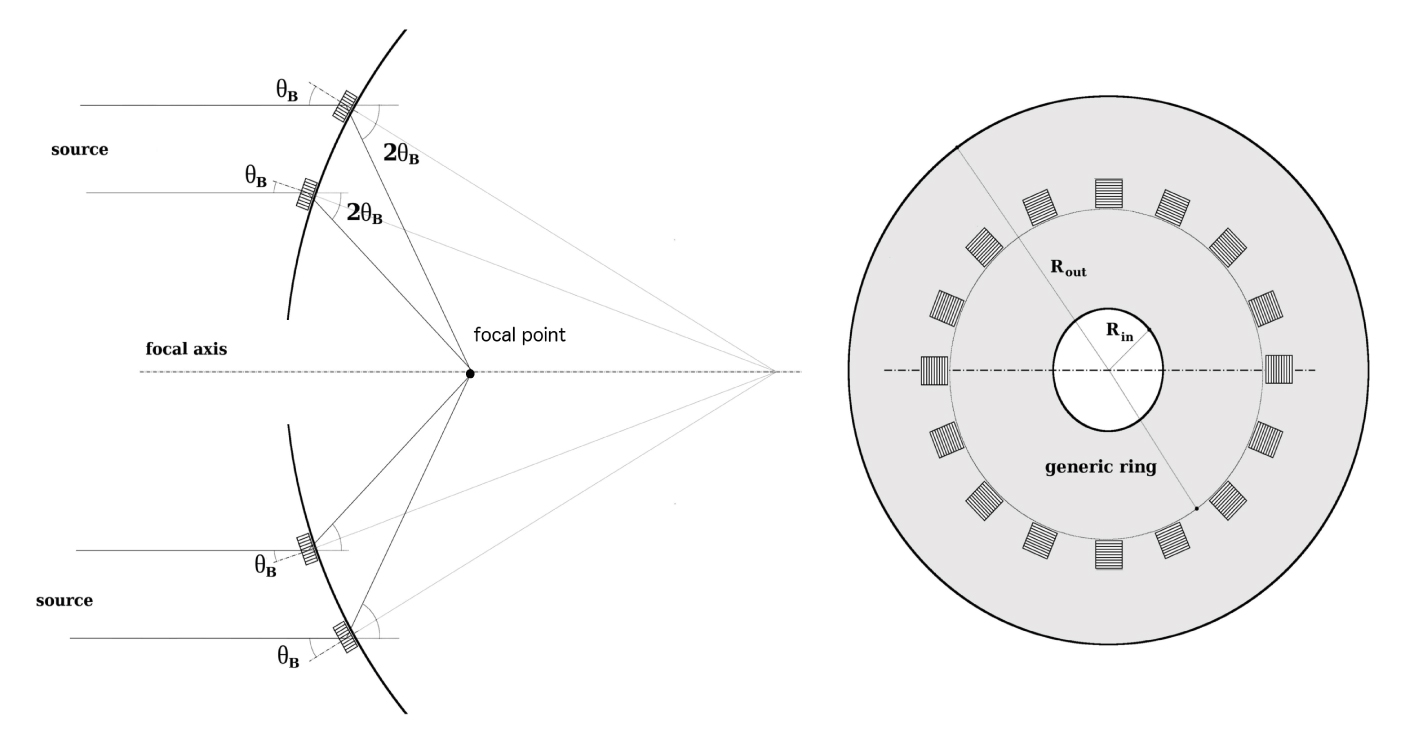}
    \caption{Side and top view of a Laue lens. The crystals are positioned on a spherical support in such a way that the radiation coming from the sky, parallel to the optical axis of the lens, interacts with the crystals and is focused. The diffraction planes of the crystals are oriented in such a way that the angle between them and the incoming X-ray beam is equal to the Bragg's angle $\theta_B$, so the angle between the diffracted beam and the incoming beam is $2\theta_B$. The focal lens is equal to half the curvature radius of the spherical cap.}
    \label{fig:laue_lens_schematics}
\end{figure}

A Laue lens, indeed, is exactly based on this concept. A Laue lens can be visualized as a spherical cap of radius $R$ covered by crystal tiles oriented in such a way that the the radiation coming from the sky is transmitted through the crystals and sent to the focal point. The focal distance of the lens $f$ is equal to half the curvature radius of the spherical cap \cite{frontera2011} (Fig. \ref{fig:laue_lens_schematics}) . 

\begin{figure}[t!]
    \centering
    \includegraphics{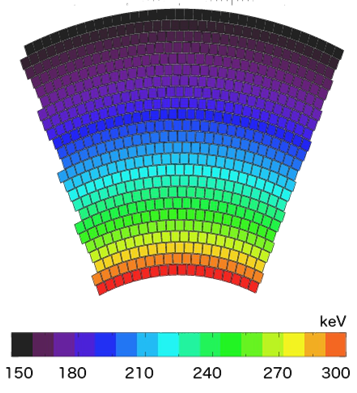}
    \caption{In a spherical Laue lens, all the crystals at the same radial distance from the center of the lens are sensitive to the same energy range, so by nesting multiple rings it is possible to build a broad-band Laue lens. The highest energies are focused by the innermost crystals, while the lower energies are focused by the outermost crystals.}
    \label{fig:laue_lens_rings}
\end{figure}

There are different configurations for the crystals on the cap which can be chosen to optimize the effective area in different ways; the simplest one, which is the configuration we are using for our prototypes, is to place the crystals in concentric rings at a distance $r$ from the axis of the lens. From the Bragg's equation, we can see that the centroid of the energy spectrum of the photons diffracted by each ring can be expressed as\cite{frontera2011}:
\begin{equation}
    E = \frac{hc}{2d_{hkl}} \sin \bigg [ \frac{1}{2} \arctan \bigg( \frac{f}{r} \bigg) \bigg] \sim \frac{hc}{d_{hkl}} \frac{f}{r}
\end{equation}
Where the approximation holds for small Bragg's angle, which is the case of hard X and Gamma rays. Higher energy photons are diffracted from crystals in the inner region of the lens, while lower energy photons interact with the outermost regions of the optics (Fig. \ref{fig:laue_lens_rings}).
With Laue lenses, we expect to be able to build a lightweight ($<$150 kg), stable optics which will enable real concentration of high energy radiation up to 600 keV at a focal distance of 20 m.
Thanks to the unprecedented imaging capabilities of Laue lenses, we will be able to obtain a Point Spread Fuction (PSF) with a width of $\sim$30 arcsec in the energy band 50 keV - 600 keV. Comparing this with one of today's most advanced high energy instruments, the IBIS aboard INTEGRAL, whose PSF is of about 12 arcmin we can immediately understand the quantum leap in imaging capabilities that would be brought by an instrument such as a Laue lens, which will also allow us to make spectroscopy and even X-ray Compton polarimetry with an unprecented quality.

\section{Bent Perfect Crystals}
\label{sect:xtals}
Several types of crystals can be used to build a Laue lens and the imaging capabilities and effective area of the optics are strictly related to the physical and chemical nature of the chosen crystals.
Two are the families of crystals that we considered for our applications: perfect crystals and mosaic crystals. 

A perfect crystal is a crystal in which the lattice plane are all oriented in the same direction. The rocking curve of the crystal has a Gaussian shape and it centered on the position satisfying Bragg's equation. The Full Width at Half Maximum (FWHM) of the rocking curve is called Darwin width \cite{Zachariasen}. A mosaic crystal, instead, can be seen as made up by microscopic perfect crystals, called {\it crystallites}, whose diffraction planes are misaligned around an average direction following a Gaussian distribution. The FWHM of the distribution of the cristallites is called {\it mosaicity} of the crystal \cite{Zachariasen}. The presence of a distribution of cristallites increases the passband of the crystals, however it also generates an effect called {\it mosaic defocusing}, which enlarges the PSF.

\begin{figure}[b!]
    \centering
    \includegraphics[scale = 0.8]{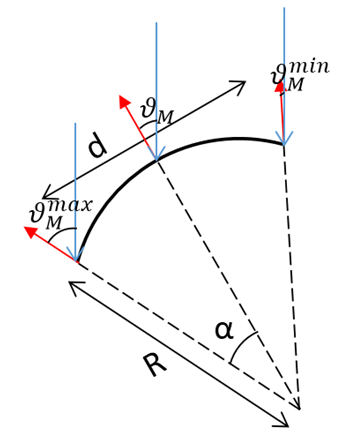}
    \caption{Explanation of the concentration effect given by cylindrical bent crystal: due to the fact that the crystal is bent, the average direction of the diffraction planes (red arrows) varies continuously inside the crystals, which means that the Bragg's angle of the incoming radiation (blue arrows) changes accordingly. The result is that the X-ray beam is concentrated and that the crystal's energy pass-band is enlarged of a quantity which depends on the length of the focusing direction $d$ of the crystal, its curvature radius $R$ and its average diffracted energy.} 
    \label{fig:curvature}
\end{figure}

With the past HAXTEL project \cite{frontera2008}, flat mosaic crystals of copper crystals were successfully used to build a prototype of a short-focal Laue lens. However they show some evident limits:
\begin{itemize}
    \item Flat crystals don't have true concentration capability: they can redirect X-ray beams, however the size of the image produced by a flat crystal is comparable to the cross section of the crystal itself. This can be a problem especially in the case of long-focal lenses, for which very small crystals would be required to obtain an adequately small image. 
    \item Flat crystals, expecially of the perfect type, show a very narrow passband\cite{Zachariasen}. Even though they can work perfectly for a narrow band Laue lens, the reduced pass-band is a problem for a broadband Laue lens, since it can be responsible of a very uneven variation of the effective area of the instrument.
    \item The diffraction efficiency for flat crystals cannot be higher than 50\%\cite{Zachariasen}. This comes from the fact that the photon traversing the crystal have the same probability of undergoing an even or odd number of diffraction processes, limiting the maximum efficiency theoretically achievable.
\end{itemize}
The use of bent crystals allow us to overcome those limitations. A bent crystal is a crystal in which the average direction of the diffraction planes varies inside the volume of the crystals according to a definite curvature radius. 
If a crystal is cilindrically bent, which is the easiest type of curvature configuration attainable, then it acts as a concentrator along the bent direction, focusing the radiation coming from an astrophysical source to a focal distance equal to half the curvature radius of the crystal itself\cite{virgilli2014}. This allows us both to shrink the cross section of the image in a region smaller than size of the crystal tile, and to increase the energy passband of each crystal (Fig. \ref{fig:curvature}). 

This holds true both for perfect and mosaic crystals, however bent perfect crystals have some interesting properties which can make them particularly interesting for a Laue lens.

Bending a perfect crystal to a chosen curvature radius induce an internal secondary curvature on specific diffraction planes in a way related to the primary curvature radius of the crystal\cite{Authier98}. In this way, the average direction of the diffraction planes varies continuously inside the crystal, slightly increasing their angular spread, whose average value is called {\it quasi-mosaicity} of the crystal \cite{Camattari2015}.

\begin{figure}[t!]
    \centering
    \includegraphics[scale = 0.6]{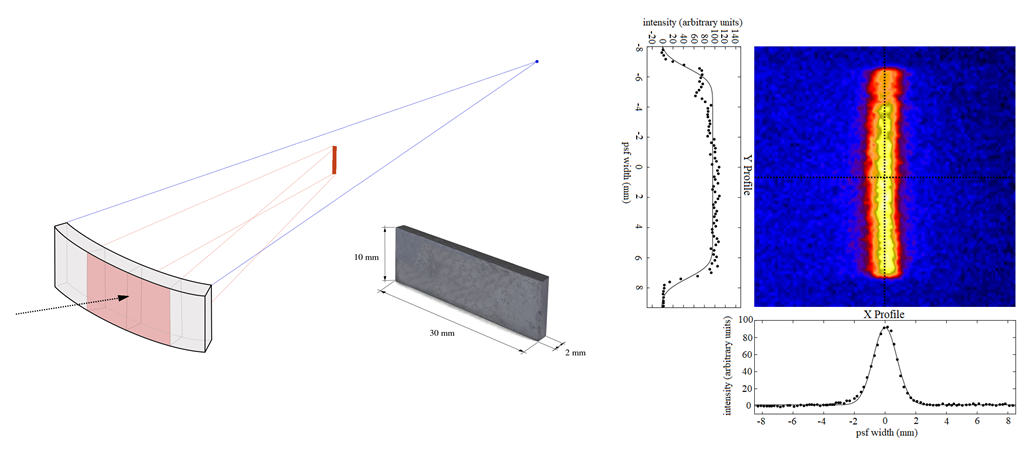}
    \caption{Right: Schematics of the concentration effect of one cylindrical bent, 30$\times$10 mm$^2$ crystals such as the one used in our tests. The radiation (in red) is concentrated along the focusing direction and the image is produced at a distance equal to half the curvature radius of the crystal. Left: Experimental image of a concentrated 150 keV X-ray beam obtained at the LARIX Facility. The area of the crystal is of 30$\times$10 cm$^2$, while the area of the image is of 0.1$\times$10 cm$^2$, which means that the beam was concentrated in an area 30 times smaller than the size of the crystal itself.}
    \label{fig:concentration}
\end{figure}

According to the dynamical theory of diffraction in bent perfect crystals, the bending of a perfect crystal can also increase the reflectivity of the bent planes above the limit of 50\%\cite{Authier98}. This can be explained intuitively in the following way: since the diffraction planes are curved, the probability that the beam crossing the crystal undergoes further diffractions gets reduced in a significant way. Experimental results suggest that an increase of the reflection efficiency can be effectively achieved\cite{Bellucci2013}.

A crystal can be curved in an elastic way, by applying mechanical clamps to its surface, or in a plastic way, by 
subjecting it to a permanent deformation \cite{Smither2005}. A mechanical bending of the crystal can bring some excessive complications for an astrophysical Laue lens, since adding further mechanical parts would increase the weight of the optics, the complexity of the design and the opacity to radiation, so crystals permanently bent are required for this type of applications. 

Different methods to obtain a self-standing, cylindrical curvature have been studied and the surface lapping technique, developed at the IMEM/CNR in Parma, is one of the methods that reached a good technological maturity 
\cite{Ferrari2013} citare Buffagni2015. This technique consists in inducing a controlled damage by lapping one of the surfaces of the crystals, which generates an internal strain able to bend the crystal in self-standing way. The parameters of the lapping process, such as the grain of the machine, the sanding time and the final thickness of the crystals after the lapping procedure, can be set to bend the crystals to the desired curvature radius. The process can be paired with an etching process which can be used to reduce an excess of the curvature radius. By combining both methods, the curvature radius of the crystal can be fine-tuned in a very precise way. The focusing capabilities and the pass-band increase of crystals bent by surface lapping have been experimentally confirmed at our facility\cite{Virgilli2016}.

For our prototype of Laue lens, we aim to use bent perfect crystals of Silicon and Germanium. Past tests done with mosaic GaAs(220) crystals curved to a radius of 40 m showed that bent crystals are able to focus high energy X-rays (Fig. \ref{fig:concentration}). Now that the focusing capabilities of bent crystals has been widely assessed, we are moving to perfect Si(111) and Ge(111). According to the dynamical theory of diffraction, the (111) planes will show the properties of increased efficiency described by Malgrange's theory of bent perfect crystals.

\section{The TRILL project}
\label{sect:trill}
\begin{figure}[b!]
    \centering
    \includegraphics[scale = 0.7]{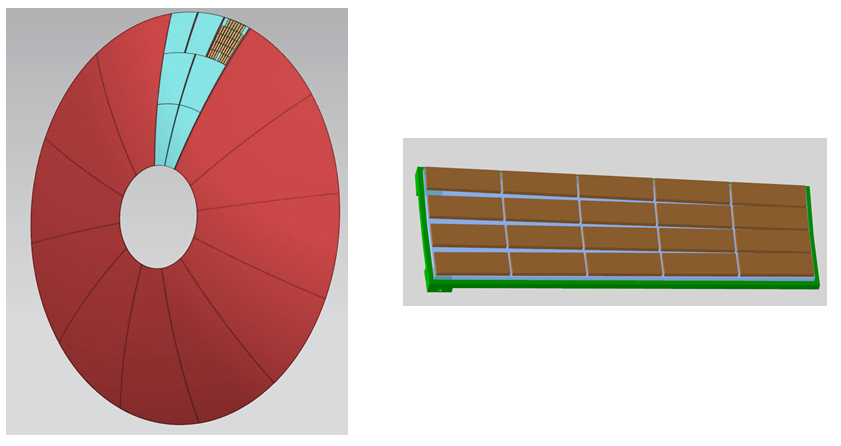}
    \caption{Left: Schematic CAD model of a full Laue lens. The lens is divided in spherical sectors, called petals (red), and each petal is divided in a series of modules (cyan). The crystals are fixed on each module. Right: CAD model of a module of Laue lens, with 20 crystals on top.}
    \label{fig:TRILL_CAD}
\end{figure}

The TRILL project (Technological Readiness Level Increase for Laue Lenses) is a project funded by the Italian Space Agency (ASI) for the increase of the Technological Readiness Level (TRL) of technologies with a potential use in future space missions. 
The aim of the TRILL project is to increase the TRL of a Laue lens in all its aspects, from the production of the crystals, to their integration on the lens itself, to the performance testing using an adequate spectroscopic-imager detectors used as focal plane detectors. 

To build our first prototype of Laue lens, we decided for a modular approach: the full Laue lens in divided in spherical sectors called {\it petals} and each petal is then divided in smaller {\it modules}. Every module can host some tenth of crystals (Fig. \ref{fig:TRILL_CAD}). Aim of the TRILL project is to build 4 modules of a Laue lens and join them together in a first prototype of a subsector of a petal.

The TRILL project is structured on the following main tasks: 
\begin{enumerate}
    \item To define a reliable and repeatable way to bend the crystals with no deterioration of their properties. To avoid distortion on the PSF of the single crystals, it is crucial that the curvature radius of the bent Si and Ge crystals is as uniform as possible and within an accuracy of $\pm 2$ m from the desired curvature radius of $40$ m.
    \item To find the best materials and bonding method to build a module of Laue lens. To reach the desired PSF size of 30 arcsec, we need that the crystal tiles are set with an accuracy better than 10 arcsec, which requires us to find a very fine way to position the crystals.
    \item To find a way to join the modules together in one single piece. The modules will be connected in a way that will grant the possibility to adjust the position of each piece in an active way, to allow us to correct the position of the modules whenever required.
\end{enumerate}
In this section, we illustrate the status of advancement of each of those 3 tasks. 

\begin{figure}[b!]
    \centering
    \includegraphics[scale = 0.6]{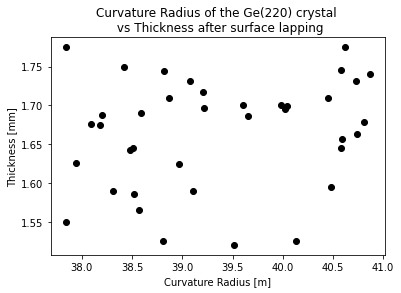}
    \caption{Curvature radius of the sample of 38 bent crystals of Ge(220) vs their thickness after the surface lapping procedure. All the curvature radius are in the limit $\pm$2 m from the nominal curvature radius of 40 m. }
    \label{fig:RvsT}
\end{figure}

The manufacturing of the crystals is done at the CNR/IMEM in Parma. Here, the crystals of Silicon and Germanium that we want to use for our prototypes are cut from wafers to obtain $30 \times 10 \times 2$ mm$^3$ tiles. Each tile is then bent via surface lapping to the desired curvature radius of $40$ m. The procedure in the past was first tested on mosaic GaAs(220) crystals with a mosaicity of 13 arcsec. From those previous tests, we discovered that there is range of values for the crystal curvature radius for which the distortion of the image induced by a distorted curvature radius is completely masked by the mosaic defocusing effect. This allows us to reduce the constraints on the accuracy of the curvature radius of the crystals according to the value of the angular dispersion of the diffractive planes.

In the case of perfect Ge(220) crystals, the lapping procedure generates a mosaic structure on the diffraction planes of the crystals, which acquire a mosaicity of $\sim$5 arcsec. Given this value of mosaicity, if the curvature radius of the crystals is within $\pm 2$ m from the nominal value of $40$ m, the induced distortion on the PSF is negligible. 

Last tests on Ge(220) gave very promising results: the values of the curvature radius of this first sample of 38 Ge(220) crystals were in the range [37.9 $\pm$ 0.9; 40.9 $\pm$ 0.7] m, which means that every crystal is suitable for our applications. The average value of the curvature radii distribution is of 39.3 $\pm$ 0.2 m, with a standard deviation of 1 m (Fig. \ref{fig:RvsT}). 


The lapping procedure, however, removes part of the material and the thickness of the crystals gets significantly decreased. The final thickness of the crystals after the lapping procedure is in the interval [1.520 $\pm$ 0.005; 1.775 $\pm$ 0.005] mm, with an average value of (1.665 $\pm$ 0.002) mm (Fig. \ref{fig:RvsT}). 


\begin{figure}[b!]
    \centering
    \includegraphics[scale = 0.6]{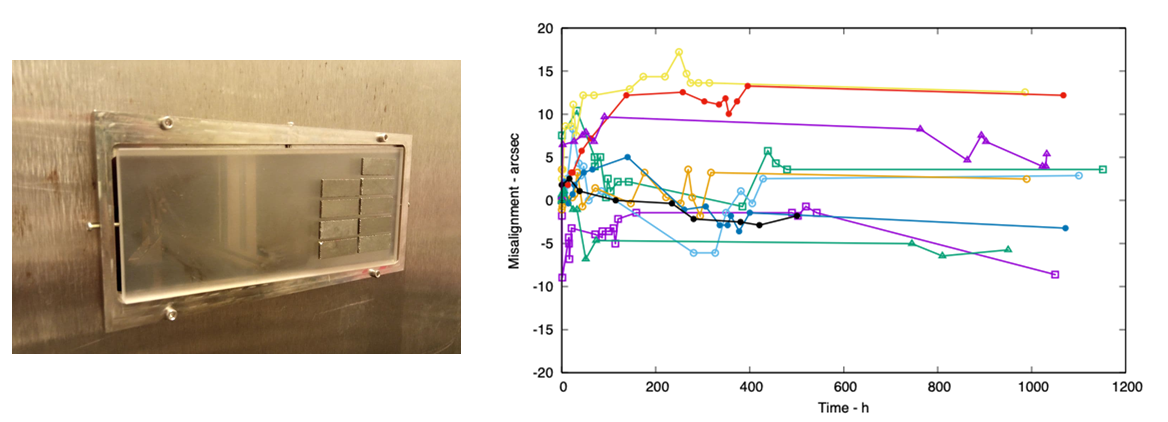}
    \caption{Left: Laue lens module with nine crystals glued on top. The steel frame supporting the glass substrate is visible. Right: Misalignment of the crystals vs time from bonding.}
    \label{fig:laue_sect}
\end{figure}

The module assembly is the second, crucial task of the TRILL project. The crystals will be bonded on an adequate substrate.  Ray-tracing simulations of a full Laue lens made by bent perfect crystals of Ge(111) allowed us to establish that it is necessary a positioning accuracy of the crystals'diffracting planes $<$10 arcsec to obtain the desired PSF accuracy of 30 arcsec for the full lens. Given that the focal length of the lens is of 20 m, this required accuracy translates in a positional accuracy of the crystals of the order of few microns, which means that high precision bonding techniques are required to integrate the crystal on the support of the prototype.

Currently each Laue lens module consists of a glass trapezoid substrate on which the crystals are glued with a low-shrinkage UV-curable glue.
Each substrate is made of clear quartz, chosen due to its transparency and its low CTE. The substrate is shaped as an isosceles trapezium with bases 68 mm and 56 mm long, while the height is 183 mm. Given that the cross section of the crystals is 30$\times$10 mm$^2$ and that the spacing between the crystals glued on the substrate will be of about 1 mm, each glass support can hold about 30 crystals. The surfaces of the substrate 
was bent to a curvature radius of 40 mm, the same of the crystals, at the National Institute of Optics of the National Research Council, in Florence. The substrate is hold in vertical position by an INVAR steel frame.

The crystals are positioned on the glass substrate using a motorized hexapod with six degrees of freedom: three translations and three rotations. The hexapod has a accuracy of the order of 1 micron on the translational movements and of 2$\times$10$^{-5}$~radians on the rotations (Fig. \ref{fig:laue_sect}, left). 

The glue currently used to bond the crystals to the glass substrate is the {\it DYMAX OP-61-LS}, an UV curable, single component optical adhesive. This glue was choosen for its very low linear shrinkage of 0.03\% of its length and for its high viscosity, which makes it suitable for a vertical bonding. 

The crystals are glued with the following procedure: (1) the crystal is positioned on the hexapod and oriented to the proper Bragg's angle, then (2) a small drop of glue is deposited on the substrate with a glue dispenser, (3) the crystal is then put in contact with the glue and the position is fine tuned to correct eventual slight change in the position of the crystal, (4) the glue is cured with an UV lamp and finally (5) the crystal is released from the hexapod. 

The position of the crystals is then tested immediately after the bonding and also when some time has passed. We found that the misalignment of the crystals after 50 days from bonding is within the interval +15/-10 arcsec (Fig. \ref{fig:laue_sect}, right). These first results are quite good, since we are starting to get close to our target of $<$10 arcsec misalignment. However at the moment this technique shows some problems of stability in time and repeatibility. In particular, we are studying a better way to measure the quantity of glue deposited on the substrate and to control the illumination uniformity of the UV light used for the curing process. 

\begin{figure}[t!]
    \centering
    \includegraphics[scale = 0.8]{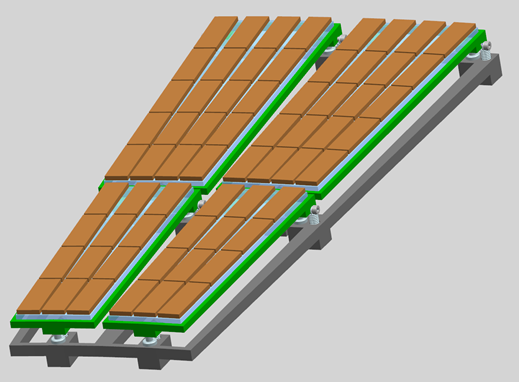}
    \caption{CAD model of the four Laue lens module assembled. The module will be connected to a steel frame by means of micrometric screws.}
    \label{fig:full_prot}
\end{figure}

The last task, i. e. the study of a way to assemble the modules together, is currently in study phase. At the moment, we plan to join the modules together by building a metal support frame in which the different modules will be fixed with high precision micrometric screws (Fig. \ref{fig:full_prot}). The screws will be set in position by using miniaturized piezoelectric motors, which will allow us to actively change the orientation of the module in any moment and to reduce as much as possible the effect that external solicitation can have on the final alignment of the prototype and, latter, of the full Laue lens itself.

The next step of the TRILL project will be to start the production of the bent Si(111) and Ge(111) crystals, to further optimize the bonding process and build the 4 modules and, finally, to assemble the modules together.

\section{The NFT aboard ASTENA}
\label{sect:nft}
\begin{figure}[t!]
    \centering
    \includegraphics[scale = 0.7]{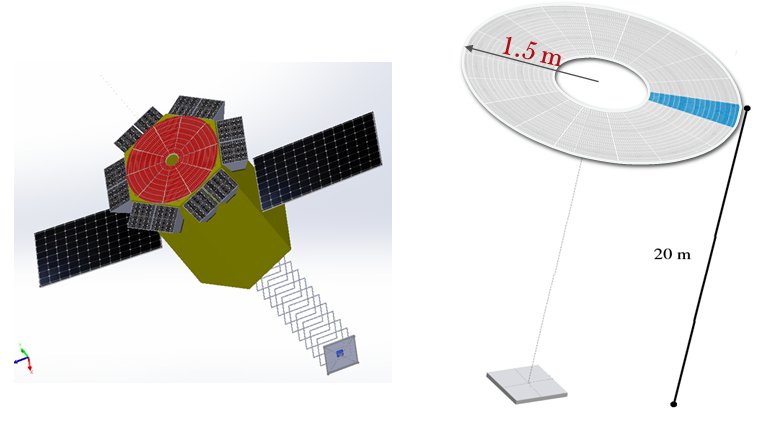}
    \caption{Left: Representation of ASTENA's in-flight configuration. The Laue lens is shown on the top of spacecraft, in red. Right: Schematic representation of NFT. NFT will be made by a Laue lens optics with 3 m of diameter and 20 m of focal length. The focal plane detector will be a CZT spectral-imager detector.}
    \label{fig:ASTENA}
\end{figure}

Our studies on Laue lenses, including the TRILL project, are oriented to develope the technology for building the Narrow Field Telescope (NFT) on-board ASTENA, the Advanced Surveyor for Transient Events and Nuclear Astrophysics, a concept mission that we proposed for the ESA Call "Voyage 2050"\cite{Frontera2021}\cite{Guidorzi2021}.
The Narrow Field Telescope will be a revolutionary hard X/soft Gamma-ray focusing telescope working in the energy band 50-700 keV based on the technology of Laue lenses (Fig. \ref{fig:ASTENA}). The NFT will be made by a Laue lens optics of 3 m of diameter with 20 m focal length. The lens will be made by about 19500 crystal tiles of perfect Si(111) and Ge(111) of size 30$\times$10$\times$2 mm$^3$, bent to a curvature radius of 40 m. In the context of the TRILL project, we will then test the capabilities of the crystals that will be used for the lens and define the construction method for the whole lens. 

The assumed focal plane detector will be a pixelated CdZnTe spectral-imager detector, with a size of 8$\times$8$\times$8 cm$^3$, a pitch of 300 $\mu$m and an efficiency $>$80\% in the whole energy band of NFT. 

The combination between the characteristics of the optics and the detector will grant NFT of an unprecedented angular resolution in the sub-MeV energy range of 30 arcsec and a point source localization accuracy $<$10 arcsec, with a Field of View of 4 arcmin. NFT will bring a leap in sensitivity of two order of magnitude with respect to the best current instruments working in the same energy bands, opening a new range of possibilities for high energy astronomical observations.

\section*{Acknowledgements}
This work is partly supported by the AHEAD-2020 Project grant agreement 871158 of the European Union’s Horizon 2020 Programme and by the ASI-INAF agreement no. 2017-14-H.O "Studies for future scientific missions".

\bibliographystyle{ws-procs961x669}
\bibliography{bibliography}

\end{document}